# Freezing-induced ordering of block copolymer micelles

Jérémy Dhainaut, Giulia Piana, Sylvain Deville,
Christian Guizard, and Michaela Klotz

Laboratoire de Synthèse et Fonctionnalisation des Céramiques, UMR3080,
CNRS/Saint-Gobain, 550 Av. Alphonse Jauffret, 84306 Cavaillon Cedex, France

**Abstract**

We demonstrate here the ordering of block copolymer micelles by ice templating, below 0 °C. We used this for the preparation of silica monoliths that present an ice-templated macroporosity, combined with a 2D hexagonal mesostructure templated by the addition of P123. We propose a mechanism triggered by the progressive freezing-induced concentration.

Monoliths that combine hierarchical porosities in the range of mesopores and macropores have attracted widespread interest in catalysis and separation processes. Surfactant self-assembly can be combined with various macropore forming methods such as foaming,[1,2] spinodal decomposition,[3] or hard templating by colloidal crystals.[4] However, the macropores formed by these routes are tortuous. Conversely, ice templating is a processing route that yields straight macropores. Ice templating takes advantage of ice crystal growth to template macropores in the range of micrometers to tens of micrometers.[5] Under unidirectional solidification conditions, the ice crystals grow preferentially along the temperature gradient direction. As the solubility limit of almost any compound in ice is extremely low, any species present in the initial suspension will be rejected and progressively concentrated by the growing ice crystals. Sublimation of the ice crystals, followed by thermal treatment, leads to centimeter-sized monoliths with directional macropores.

Extensive research on silica thin film formation by evaporation induced self-assembly (EISA) showed that well-ordered mesoporous materials can only be obtained by controlling the hydrolysis and condensation of silicon alkoxide.[6,7] Building upon this experience, we want to trigger such concentration induced self-assembly by removing the solvent by freezing instead of evaporation. Similar sol–gel parameters (pH, ageing temperature, time, etc.) may then be relevant for structural control by ice templating. However, the process temperatures are below 0 °C, where, notwithstanding rare exceptions,[8] surfactants do not self-assemble into lyotropic liquid crystal mesophases. Pluronic 123 (P123) is a triblock copolymer composed of ethylene oxide (EO) and propylene oxide (PO) blocks with general formula $(EO)20(PO)70(EO)20$. The solubility of its EO and PO blocks is sensitive to temperature, and spherical micelles are only formed between 20 °C and 65 °C, even under acidic conditions.[9] Nonetheless, we demonstrate here that a favorable combination of ice templating, surfactant self-assembly, and sol–gel chemistry allows mesophase ordering of P123 by ice templating.

P123 was first dissolved in water at a mild temperature (<35 °C). In order to favor hydrolysis and limit condensation, the pH of the water was adjusted between 1.1 and 1.8 using hydrochloric acid. Tetraethylorthosilicate (TEOS) was then added to reach the following molar compositions: 1 $SiO_2$ : 71–83 $H_2O$ : 0.017 P123. After ageing (<35 °C, 3 h), the as-formed ethanol was removed by rotary evaporation to increase the freezing point of the suspension. This suspension was then divided into two equivalent parts. The first one was let to evaporate at 90 °C in an oven, to trigger EISA. The second part was ice-templated, following the procedure described in S1 (ESI†), to cause freezing-induced self-assembly (FISA). Freezing of the solution occurred at −15 °C. After freeze-drying to selectively remove the ice crystals, both samples were thermally treated at 450 °C for 4 h with a heating rate of 1 °C min−1.

The mesostructure of EISA and freezing-induced self-assembled (FISA) samples were explored by TEM (Fig. 1 and Fig. S2, ESI†). 2D hexagonal structures with long-range ordering are observed in both cases. The pore-center-to-pore-center average distance was estimated to be 10.9 nm. Small angle X-ray diffraction patterns (Fig. 2 and Fig. S3, ESI†) of both samples exhibit three reflections that can be indexed according to the p6mm symmetry, the 2D hexagonal structure. The inter-reticular distances of reflections in both patterns are comparable. The most intense peak, at $2\theta = 0.96°$, can be indexed to the (10) plane, while reflections at 1.67° and 1.93° correspond, respectively, to the (11) and (20) planes of the planar p6mm space group. The corresponding cell parameter is 10.6 nm, a value consistent with the pore-center-to-pore-center distance estimated by TEM. These results clearly show that FISA, under the conditions reported here, results in organized mesoporosity. The diffracted intensities measured are a combination of the form factor and the structure factor. In the case of P123, both responses are in the same scattering region. This impacts the total intensity and can modify relative intensities between the diffracted peaks.10 TEM analysis of both samples indicates similar quality of ordering for EISA and FISA.

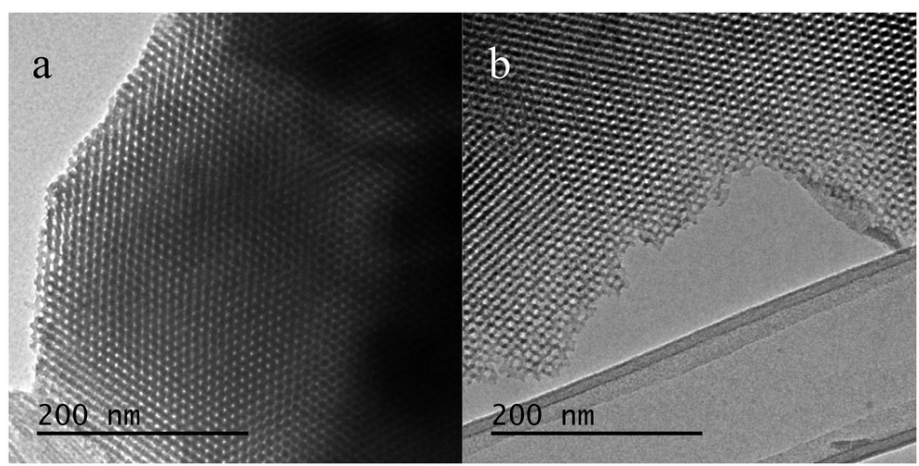

Figure 1: Comparative TEM images of (a) EISA and (b) FISA-templated mesoporous silica.

The macroscopic pore structure obtained by FISA is shown in Fig. 3A. The arrow indicates the freezing direction. In comparison with macropores typically obtained for ice templated suspensions,11 the macropore structure has an open, foam-like morphology. A closer view of the fracture surface (Fig. 3B) reveals a 1.3 μm thick secondary pore structure, perpendicular to the macropore. TEM observations (Fig. S4, ESI†) confirmed the bundle-like organization in this region. Ordered mesopores obtained from P123 templating are oriented parallel to these bundles. This orientation is lost halfway through the inorganic wall. The interface between oriented and disordered pores is clearly visible. In the inner region, TEM observations (Fig. S5, ESI†) reveal a more random orientation of the mesopore domains.

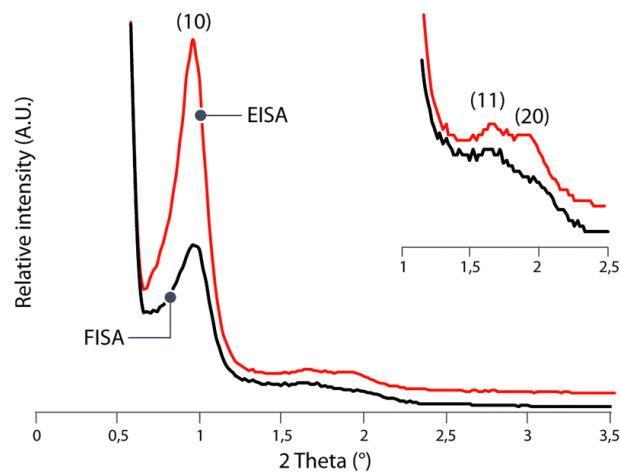

Figure 2: Comparative small angle X-ray diffraction patterns of EISA and FISA-templated mesoporous silica.

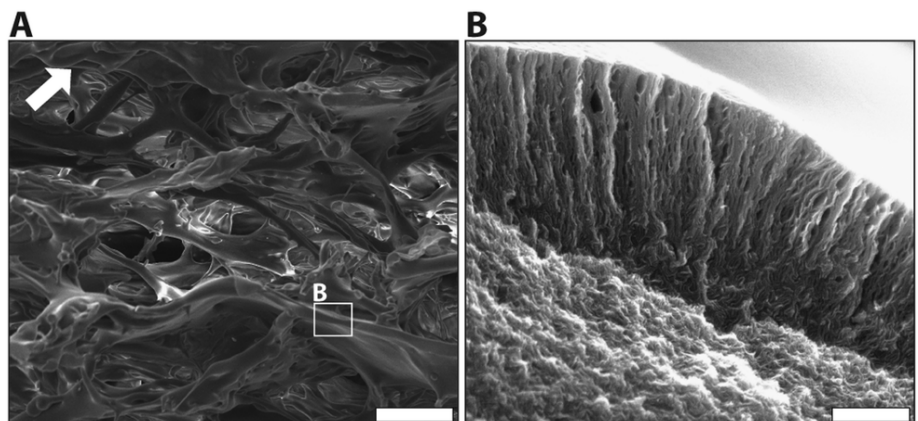

Figure 3: SEM images of FISA-templated mesoporous silica. Scale bars: 100 µm (A) and 500 nm (B).

Classical type IV nitrogen adsorption–desorption isotherms, characteristic of mesoporous materials, with H1-type hysteresis that ends in the H4-type are observed (Fig. 4 and Fig. S6, ESI†). H4 hystereses are typical of interlayer porosity and characteristic here of the bundle-like organization. Pore size distribution and microstructural observations reveal several levels of interconnected porosity: ice-templated macropores (40–80 µm and 150–190 µm, Fig. S7, ESI†), bundle-like macropores (50–200 nm, Fig. S4, ESI†), and a hexagonal ordered mesoporosity (5.5–7.8 nm). The mesopore characteristics can be adjusted by variations of the sol–gel (pH, ageing time) and freezing (cooling rate) parameters (Table S2 and Fig. S6, ESI†).

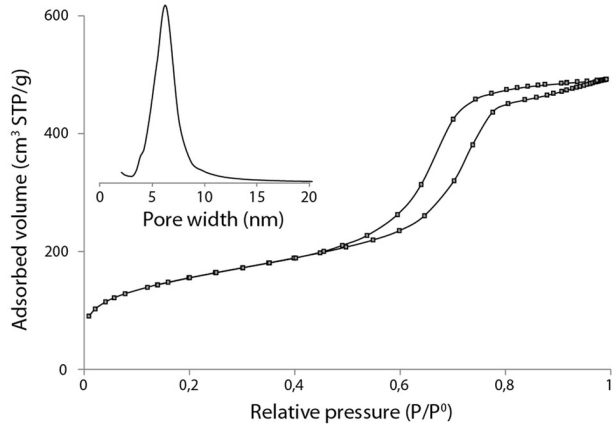

Figure 4: Nitrogen adsorption–desorption isotherm of FISA-templated mesoporous silica. The inset represents the BJH mesopore size distribution.

During ice templating, water transforms into ice crystals, which results in the segregation of any components, i.e. silica oligomers and tri block copolymer micelles, initially incorporated into the solution. This concentration effect is significant, as concentration will increase until a maximum value is reached.[12] HCl will also be rejected and locally shift the pH. Based on the fraction of ice crystals, and assuming that all species are rejected from the growing crystals, we can estimate the pH to decrease from 1.3 to 1. Therefore we propose, from the aforementioned experimental results, a mechanism in which the silica oligomers and the micelles are rejected by the ice crystals and rearrange themselves to form an ordered mesoporous structure. The micelles serve as sacrificial templates for mesopores whereas ice crystals are responsible for creating macroporous skeleton walls at the upper level. Interestingly, the SiO2/surfactant ratio (0.017) at which ordering occurs upon freezing is identical to the usual SiO2/surfactant ratio for the synthesis of SBA-15.[13] Thus, in comparison with EISA,[14] we can also speculate that the extraction of the solvent by freezing induces block copolymer concentration, leading to an ordered mesostructure. However, at this stage, there is no experimental evidence of this mechanism and of whether the micelles may pre-exist in the initial solution or can be formed upon freezing. More experimental data are needed to identify the formation mechanism, which could be self-assembly, crystallisation, or precipitation. Due to the dynamic nature of the process, there is interplay between the thermodynamic and kinetic conditions, the question of whether the thermodynamic equilibrium is achieved or not is thus still open. The local conditions at the freezing front eventually change due to concentration by solvent extraction. Moreover they are probably different from the batch solution as for example the latent heat released by the solidification of water will impact the local temperature. Considering the experimental data of the present work, it can be only argued that the increase of the concentration triggers ordering of the micelles, which results in an ordered mesoporous structure.

The technique described here for the preparation of macroporous silica monoliths with organized mesoporous walls is reproducible (Tables S1 and S2, ESI†) and could potentially be suitable for the fabrication of other multi-scaled porous inorganic materials. These results are of interest at several levels. From a materials science point of view, the complex hierarchical porous materials exhibit a unique combination of very high pore volume, well-calibrated mesopores, and oriented macropores. Such structures may be of interest for applications where diffusion kinetics are critical, such as adsorbents, catalysts, and catalyst carriers, especially in liquid media. Because FISA occurs in the bulk, centimetre-sized or even larger monoliths can be processed. The principles of FISA described here are generic and may be extended to other self-assembly systems, inorganic matters, and solvents, relying on concentration gradients to trigger their organization.

## Acknowledgements

Financial support from Saint-Gobain is gratefully acknowledged. A. Addad is acknowledged for help with TEM imaging. The TEM facility in Lille (France) is supported by the Conseil Regional du Nord-Pas de Calais, and the European Regional Development Fund (ERDF).

## Notes and references

**Footnote**

† Electronic supplementary information (ESI) available: Detailed synthesis protocol, small angle X-ray diffraction patterns, TEM images and mercury porosimetry. See DOI: 10.1039/c4cc05556j